\newcommand{\vc}[1]{\ensuremath{\mathbf{#1}}}

\newcommand{\ket}[1]{\ensuremath{\left|  #1 \right\rangle}}
\newcommand{\Rb}{\ensuremath{^{87}\mathrm{Rb}}}

\documentclass[aps,pra,reprint,showpacs,amsmath,amssymb]{revtex4-1}
\usepackage{graphicx}
\usepackage{dsfont}
\usepackage{hyperref}
\begin{document}
\title{General Formalism for Evaluating the Impact of Phase Noise on Bloch Vector Rotations}
\author{Zilong Chen}
\email{zchen@jila.colorado.edu}
\author{Justin G. Bohnet}
\author{Joshua M. Weiner}
\author{James K. Thompson}
\affiliation{JILA, NIST, and Department of Physics, University of Colorado, Boulder, Colorado 80309-0440, USA}
\date{\today}

\begin{abstract}
Quantum manipulation protocols for quantum sensors and quantum computation often require many single qubit rotations. However, the impact of phase noise in the field that performs the qubit rotations is often neglected or treated only for special cases.  We present a general framework for calculating the impact of phase noise on the state of a qubit, as described by its equivalent Bloch vector. The analysis applies to any Bloch vector orientation, and any rotation axis azimuthal angle for both a single pulse, and pulse sequences. Experimental examples are presented for several special cases.  We apply the analysis to commonly used composite $\pi$-pulse sequences: CORPSE, SCROFULOUS, and BB1, used to suppress static amplitude and detuning errors, and also to spin echo sequences.  We expect the formalism presented will help guide the development and evaluation of future quantum manipulation protocols.
\end{abstract}

\pacs{03.67.-a, 05.40.Ca, 76.60.Lz, 82.56.Jn}
%\keywords{Suggested keywords}%Use showkeys class option if keyword
%display desired
\maketitle

\section{Introduction}
Atomic quantum sensors and tests of fundamental physics commonly rely on the ability to rotate a Bloch vector representing a spin-$\frac{1}{2}$ system or qubit. Besides quantum state manipulation, rotations can also be used to undo inhomogeneous errors or to reduce other sources of noise, as is done with spin echo pulses or dynamical decoupling~\cite{CP54, MG58, VKL99, Uhrig07, MUV09a, MUV09b, Koschorreck10b}.  Precise rotations, required for manipulating collective spin-squeezed states~\cite{Appel09, Schleier-Smith10, LSV10, GZN10, RBL10, CBS11, Hamley12, Louchet-Chauvet10, LMV10}, might also enable dynamical spin-squeezing to the two-axis squeezing limit~\cite{LXJ11} using already realized one-axis twisting in cold atoms systems~\cite{LSV10, GZN10, RBL10, Hamley12}.  Given that actual rotations are imperfect, it is an open question whether such rotation protocols can be realized without adding large amounts of additional noise, thus destroying the squeezing.

Most rotation protocols assume that the phase of the field that rotates the qubit is perfectly stable, and that imperfections arise only due to slowly varying amplitude errors or detuning errors.   Composite rotations sequences~\cite{TCS85, Wimperis94, CLJ03, BHC04, RLL09} and generalizations to shaped pulses~\cite{Steffen07, Torosov11}, including optimal control theory~\cite{Khaneja05, Li06, Timoney08}, can be used to reduce these errors to essentially arbitrary order.

In reality, the phase of the qubit-field coupling is never perfectly stable, largely due to phase noise in the local oscillator (LO) used to generate the field.  The LO is typically a radio or microwave oscillator in nuclear spin, superconducting Josephson junction, quantum dot, neutral atom/ion Zeeman and hyperfine qubit systems~\cite{MI00}.  In the case of highly-forbidden optical transitions~\cite{YKK08}, the LO is an ultra-stable laser. Further, qubit transition frequency fluctuations can be straightforwardly mapped onto an equivalent phase noise of the LO.  Such a fluctuation might arise due to noise in the  DC bias current  of a superconducting Josephson junction qubit~\cite{MNA03}, or differential light shifts for atomic qubits in an optical trap~\cite{YKK08}.   It is critical for future work beyond proof-of-principle experiments to develop general tools for analyzing the impact of phase noise on a rotation, both in terms of overall fidelity for quantum gates, and quadrature specific noise for manipulating states with anisotropic sensitivity to noise, such as spin-squeezed states~\cite{Appel09, Schleier-Smith10, LSV10, GZN10, RBL10, CBS11, Hamley12, Louchet-Chauvet10, LMV10}, reduced spin-noise states~\cite{Takano10}, and Dicke states~\cite{Dicke54}.

The effect of phase noise on atomic response has been studied in various contexts, for example on atomic excitation probability~\cite{AC77, ZE77, EWS84, Camparo02} under continuous drive.   In the atomic sensor community,  the impact of phase noise is evaluated for a single quadrature for specific Ramsey sequences~\cite{Dick87, SAM98, CCP08}.  In contrast, we present a general framework that may be applied to an arbitrary LO  phase noise spectrum for continuous resonant drive without making assumptions about the orientation of the Bloch vector. The framework can be extended to arbitrary resonant pulse sequences assuming white LO phase noise. Furthermore, we fully specify all second order noise moments of the Bloch vector including covariances and variances, important for predicting the fidelity of single qubit gates and manipulations of spin-squeezed or Dicke states. The methodology and tools presented in this paper can help guide the development and evaluation of future quantum control and measurement protocols.

The organization of the paper is as follows. In Sec.~\ref{sec:deflection}, we describe the qubit-field interaction as a rotation of a Bloch vector. We show that the net effect of a coherently phase modulated rotation can be reduced to a small rigid rotation of the Bloch sphere. A description of the experimental system used to demonstrate the theory is provided in Sec.~\ref{sec:experiment}. Experimental examples of the response of the Bloch vector driven by phase modulated rotations are then presented. In Sec.~\ref{sec:NoiseSingleRot}, we extend the analysis in Sec.~\ref{sec:deflection} assuming linear response to relate the single sideband (SSB) LO phase noise to Bloch vector noise projections through a covariance transfer matrix for a single rotation. Experimental realizations for a few special cases are also presented. The Bloch vector noise projection variances and covariances are captured in the covariance noise matrix.   In Sec.~\ref{sec:NoiseMultRot}, we generalize the single-rotation covariance noise matrix to that for multiple rotations/pulse sequences and apply the tools to commonly used composite $\pi$-pulse sequences: CORPSE, SCROFULOUS, and BB1~\cite{TCS85, Wimperis94, CLJ03, BHC04, RLL09}, and also to spin echo pulse sequences.  A simple formula for the average infidelity of any pulse sequence, appropriate in the context of quantum computing, is also presented.  Finally, we give a summary of the results presented in this paper, and an outlook of future work in Sec.~\ref{sec:conclusion}.

\section{Deflection of Bloch Vector due to coherently phase modulated rotation}\label{sec:deflection}

\subsection{System Hamiltonian}
The system considered in this paper consists of a qubit with transition frequency $f_a$ coupled to a resonant classical electromagnetic field with strength characterized by the Rabi frequency $f_R$ proportional to the field ampliutude. The system's Hamiltonian in the lab frame is
\begin{equation}
H_{\mathrm{lab}} = h f_a \sigma_z + h f_R \cos\left(2\pi f_a t + \phi(t) \right) \sigma_x \, ,
\label{eq:HamLabFrame}
\end{equation}
where $h$ is the Planck constant, $\phi(t)$ is the LO phase as a function of time $t$, and $\sigma_x\, , \sigma_y, \, \sigma_z$ are the Pauli matrices. The first term corresponds to the Hamiltonian of the qubit in the absence of the field, and the second term corresponds to the qubit-field interaction that drives Rabi flopping between the qubit states $\ket{\uparrow}$ and $\ket{\downarrow}$. By going into a rotating frame at the qubit transition frequency, and making the rotating wave approximation, the dynamics is described by the Hamiltonian
\begin{equation}
H_{\mathrm{rot}} = \frac{h f_R}{2} \left( \cos(\phi(t)) \sigma_x + \sin(\phi(t) \right) \sigma_y) \, .
\label{eq:HamRotFrame}
\end{equation}

\subsection{LO Phase Noise}
The LO phase $\phi(t)$ is random as a function of time, and its statistical properties can be captured by the autocorrelation function $\langle \phi(t) \phi(t+\tau) \rangle_t$ in the time domain, or equivalently the power spectral density of phase fluctuations $S_\phi(f_m)$ in the frequency domain, where $f_m$ is the frequency offset from the LO carrier frequency $f_{LO}$. The power spectral density of phase fluctuations $S_\phi(f_m)$ is related to the autocorrelation function through the Wiener-Khinchin theorem (see Appendix~\ref{sec:AppendixA}). The SSB phase noise $\mathcal{L}(f_m) = S_\phi(f_m)/2$ of a LO is usually specified in manufacturer datasheets instead of $S_\phi(f_m)$. To facilitate application of our results in an experimental context, the SSB phase noise $\mathcal{L}(f_m)$ is used throughout this paper. 

The phase noise spectrum of a LO is typically parametrized as a sum of  $1/f$ type phase noise as follows
\begin{equation}
\mathcal{L}(f_m) = \sum_{k = -\infty}^{\infty}  \frac{\mathcal{L}_k}{f_m^k} \, ,
\end{equation}
where the coefficient $\mathcal{L}_k$ characterizes the strength of the noise spectrum with dependence $1/f_m^k$. An important example is white phase noise where there is no frequency dependence. In this case, the only non-zero coefficient is for $k=0$ corresponding to $\mathcal{L}(f_m) = \mathcal{L}_\circ$, and the autocorrelation function is given by $\langle \phi(t) \phi(t+\tau) \rangle_t = \mathcal{L}_\circ \delta(\tau)$ where $\delta(\tau)$ is the delta function. Other frequently encountered phase noise spectra include flicker noise $\mathcal{L}(f_m) = 1/f_m$ and phase diffusion noise $\mathcal{L}(f_m) = 1/f_m^2$.

\subsection{Bloch Sphere Picture}
\begin{figure}
\includegraphics[width=3.4in]{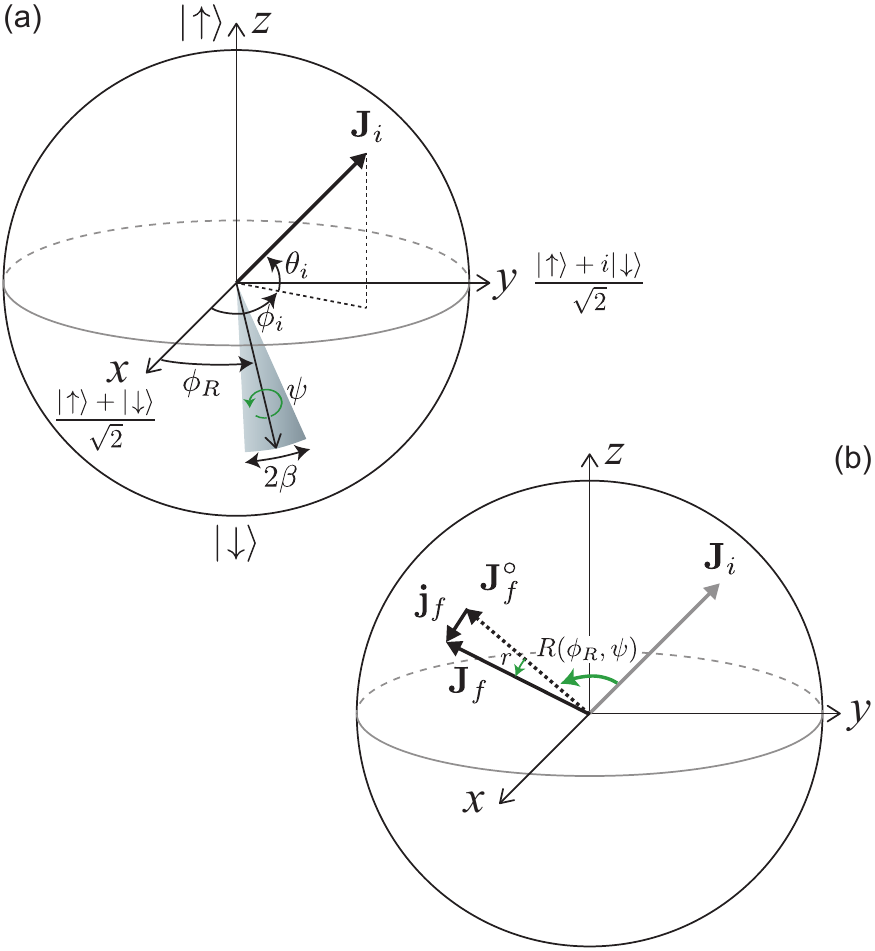}
\caption{(color online). \textbf{Bloch Sphere Picture.} (a) Geometrical representation of the Bloch vector and a coherently phase modulated rotation on the Bloch sphere.  Initial Bloch vector $\vc{J}_i$ is specified by polar angle $\theta_i$ and azimuthal angle $\phi_i$.  In the absence of phase modulation, the rotation axis is at a constant angle $\phi_R$ from the $x$-axis. With phase modulation, the rotation axis oscillates in the $x-y$ plane with amplitude $\beta$ around its average position. (b) Effect of a coherently phase modulated rotation. An unmodulated rotation $R(\phi_R, \psi)$ rotates $\vc{J}_i$ to $\vc{J}_f^\circ$. With phase modulation of the rotation axis, the final Bloch vector $\vc{J}_f$ is deflected slightly by the vector $\vc{j}_f$ that can be regarded as arising from a small rotation $r$.}
\label{fig:BlochSphere}
\end{figure}

The evolution of the qubit's state in the rotating frame is conveniently visualized using the Bloch sphere picture.  A qubit's state can be mapped onto a Bloch column vector $\vc{J}$ with components $J_k \equiv \langle \sigma_k \rangle$, where $k = \{x,y,z\}$. The tip of the Bloch vector resides on a Bloch sphere with radius equal to 1 (see Fig.~\ref{fig:BlochSphere}a).  The LO phase $\phi(t)$ sets the instantaneous rotation axis around which the Bloch vector rigidly rotates. For fixed LO phase $\phi(t) = \phi_R$, an initial Bloch vector $\vc{J}_i$ is mapped to an ideal final Bloch vector  $\vc{J}_f ^\circ$ as  
\begin{equation}
\vc{J}_f ^\circ= R(\phi_R, \psi) \vc{J}_i \, , 
\end{equation}
where $R(\phi_R, \psi)$ performs a counterclockwise rotation  through an angle $\psi \geq 0$ about an axis in the $x-y$ plane with azimuthal angle $\phi_R$ measured relative to $\vc{\hat{x}}$. The rotation angle $\psi= 2 \pi f_R t$ is set by the Rabi frequency $f_R$ and the amount of time $t$ the field is applied.

\subsection{Small Rotation describing Phase Modulated Rotation} 
Our goal is to generate a general covariance transfer matrix that maps LO phase noise onto noise projections of the final Bloch vector. To do so, we consider here the response of the Bloch vector to a coherent phase modulation of the LO phase in the frequency domain, and generalize to a phase noise process in Sec.~\ref{sec:NoiseSingleRot}. A sinusoidal modulation of the LO phase $\phi(t) = \phi_R + \beta \sin(2\pi f_m t + \alpha_m)$ represents an oscillation of the instantaneous rotation axis in the $x-y$ plane with amplitude $\beta$ about its average azimuthal angle $\phi_R$ as illustrated in Fig.~\ref{fig:BlochSphere}a. The modulation frequency and phase is $f_m$ and $\alpha_m$ respectively.

A phase modulated rotation causes a deflection of the final vector $\vc{J}_f$ from its ideal final orientation without modulation by  
\begin{equation}
\vc{j}_f (\phi_R, \psi, \beta, f_m, \alpha_m) = \vc{J}_f - \vc{J}_f^\circ \, .
\end{equation}
To first order in the phase modulation amplitude $\beta \ll 1$, the deflection is perpendicular to $\vc{J}_f^{\circ}$.    The total effect of the modulation can be described by an additional small rotation $r(\phi_R, \psi, \beta, f_m, \alpha_m)$ applied after the ideal rotation such that  
\begin{equation}
\vc{J}_f=  r \vc{J}^\circ_f \, ,
\end{equation} where labels on $r$ have been suppressed. The relationships between the vectors $\vc{J}_i$, $\vc{J}^\circ_f$, $\vc{J}_f$, $\vc{j}_f$, and the rotations $R(\phi_R, \psi)$, $r$ are depicted in Fig.~\ref{fig:BlochSphere}b.  

Because the dynamics with modulation can be described in terms of infinitesimal rigid rotations, the whole Bloch sphere is rigidly rotated by this additional phase modulation contribution, and $r$ does not depend on $\vc{J}_i$. This is confirmed by solving for $r$ analytically. Simple rotations about $\vc{\hat{z}}$ through angle $\phi_R$ relate the small rotation $r$ evaluated at $\phi_R=0$ to the small rotation evaluated at arbitrary $\phi_R$ through 
\begin{equation}
r(\phi_R) = R_z(\phi_R) r(0) R_z(-\phi_R) \, .
\label{eq:r(phi)}
\end{equation}
Therefore, it is sufficient to determine $r$ for the special case $\phi_R =0$ and $\vc{J}_i= \vc{\hat{x}}$.  Hereafter, quantities with a tilde ($\sim$) overhead apply to this special case only. We find
\begin{equation}
r(0) = R_y(-\widetilde{j}_z) R_z(\widetilde{j}_y) = 
\begin{pmatrix}
1   &  -\widetilde{j}_y & -\widetilde{j}_z \\
\widetilde{j}_y &  1    &  0 \\
\widetilde{j}_z &  0    & 1
\end{pmatrix} \, ,
\label{eq:r(0)}
\end{equation}
where $\widetilde{\vc{j}}_f = (0, \widetilde{j}_y, \widetilde{j}_z)$, and only first order in the small quantities $\widetilde{j}_y$ and $\widetilde{j}_z$ are retained.

\subsection{Solution for Deflection Vector}
We now solve for the $\widetilde{j}_y$ and $\widetilde{j}_z$ that determine the small rotation matrix $r$ defined by Eq.~\ref{eq:r(phi)} and \ref{eq:r(0)}. Writing the Heisenberg equations of motion for the Bloch vector components yields to first order in the small modulation amplitude
\begin{equation}
\frac{d\widetilde{j}_\perp}{d\psi} + \imath \, \widetilde{j}_\perp = -  \beta \sin\left(x \psi + \alpha_m \right)  \, ,
\label{eq:jpdot}
\end{equation}
where  $\widetilde{j}_\perp \equiv \widetilde{j}_z + \imath \,  \widetilde{j}_y$, and  $x = f_m/f_R$.  The rotating wave approximation $f_m \ll f_a$ has been made. Note that this is the equation of motion for the coupled position and momentum of an undamped simple harmonic oscillator with natural resonance frequency $f_R$ driven with an externally applied force at frequency $f_m$. The harmonic oscillator's displacement and velocity map onto $\widetilde{j}_y$ and $-\widetilde{j}_z$ respectively. Solving for $\widetilde{j}_y$ and $\widetilde{j}_z$ using the initial condition $\widetilde{j}_\perp(\psi = 0) = 0$, we obtain
\begin{widetext}
\begin{align}
\widetilde{j}_y &=  \frac{\beta}{1-x^2} \times \Bigl[  - x \cos\alpha_m \sin\psi - \sin\alpha_m  \cos\psi 
	+  \sin\left(x \psi + \alpha_m  \right) \Bigr] \label{eq:jy} \, , \\
\widetilde{j}_z &=  \frac{\beta}{1-x^2} \times \Bigl[  x \cos\alpha_m  \cos\psi  - \sin\alpha_m \sin\psi - x \cos\left(x \psi + \alpha_m  \right) \Bigr] \, . \label{eq:jz}
\end{align}
\end{widetext}

The above solutions can be understood as a superposition of the ``transient" and steady state response of a driven harmonic oscillator. The terms proportional to $\sin\psi$ and $\cos\psi$ in Eq.~\ref{eq:jy} and \ref{eq:jz} correspond to the response of the harmonic oscillator at its natural frequency $f_R$. This response is called the transient response in damped harmonic oscillator systems. Because there is no damping in this oscillator, the ``transient" response does not decay away. The terms proportional to $\sin\left(x \psi + \alpha_m  \right)$ and $\cos\left(x \psi + \alpha_m  \right)$ correspond to the steady state response of the oscillator at the drive frequency $f_m$. At $x=1$, corresponding to the case of driving on resonance, the solutions take on the following limits
\begin{align}
\lim_{x \to 1} \widetilde{j}_y &=  - \frac{1}{2} \left(\psi \cos\left(\psi + \alpha_m  \right) -\cos\alpha_m \sin\psi\right) \label{eq:jy_res} \, , \\
\lim_{x \to 1} \widetilde{j}_z &=  - \frac{1}{2} \left(\psi \sin\left(\psi + \alpha_m  \right) +\sin\alpha_m \sin\psi  \right) \, . \label{eq:jz_res}
\end{align}
The amplitude of the response grows roughly linearly with $\psi$ for large $\psi \gg 1$ as the drive is phase coherently adding momentum to the oscillator.

\section{Experimental Demonstration}\label{sec:experiment}
In order to connect the theory described in Sec.~\ref{sec:deflection} to an actual physical system, we experimentally demonstrate the linear response of the Bloch vector to a coherently phase modulated rotation for a few special cases.

\subsection{Physical Implementation}
\begin{figure}
\includegraphics[width=3.4in]{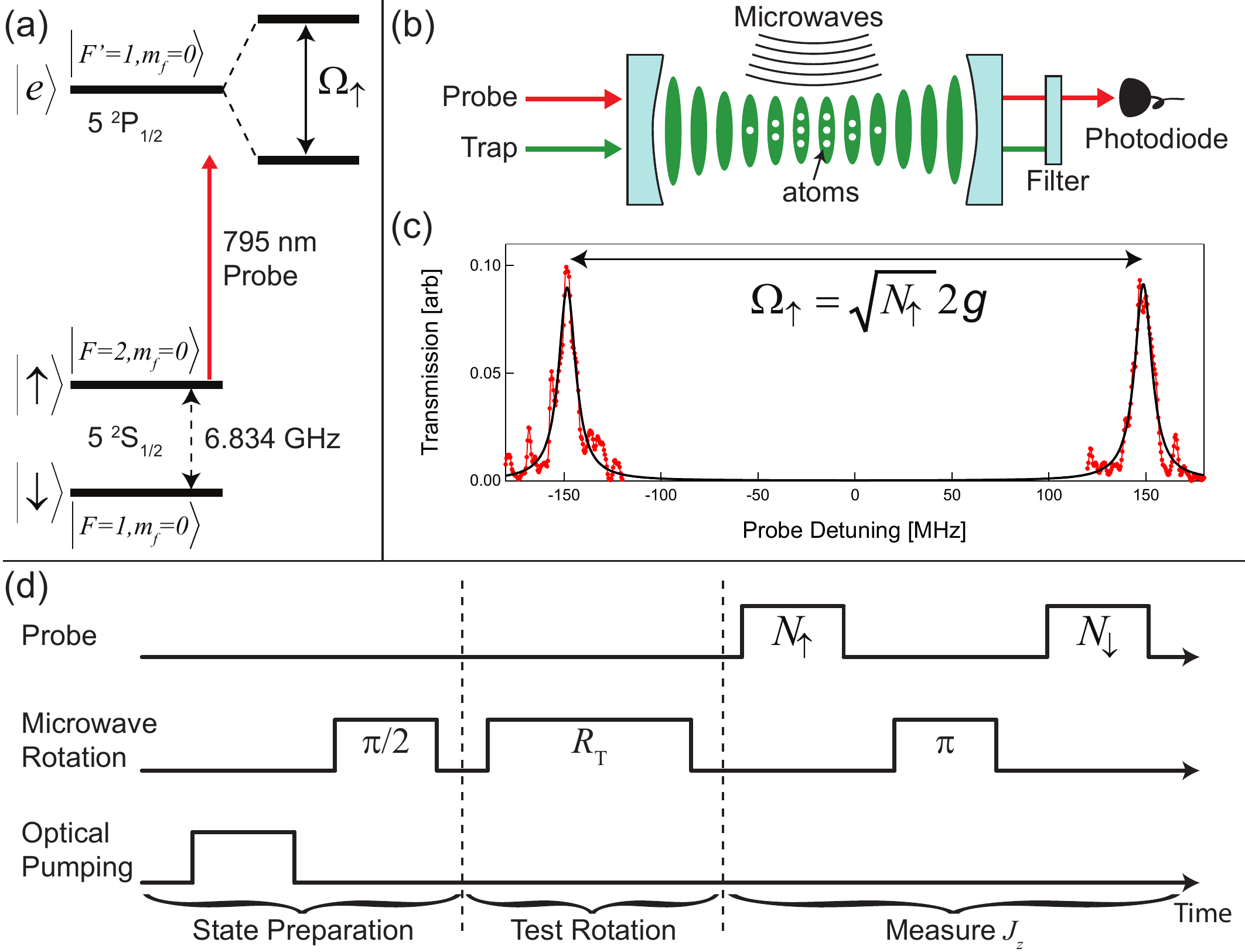}
\caption{(color online). \textbf{Experiment Schematic.} (a)  The clock states \ket{\uparrow} and \ket{\downarrow} of $\Rb$ form a pseudo spin-$\frac{1}{2}$ system. Coupling of the atoms to a cavity mode resonant with the \ket{\uparrow} to \ket{e} transition creates a collective vacuum Rabi splitting $\Omega_{\uparrow}$ which is probed to deduce the population $N_\uparrow$. (b) An ensemble of $7 \times 10^5$ atoms are trapped within the $\text{TEM}_{00}$ mode of an optical cavity using a 1D intra-cavity optical lattice at 823~nm. Rotations of the Bloch vector are accomplished using resonant microwaves at $f_a = 6.834$~GHz. The photodiode records cavity transmission as the probe laser frequency is swept across the splitting.  (c) The population $N_\uparrow =( \Omega_\uparrow/2 g)^2$ is determined from the splitting $\Omega_{\uparrow}$ obtained by fitting the cavity transmission versus probe detuning to Lorentzians. (d) Experimental sequence. After preparing the initial Bloch vector $\vc{J}_i$ on the equator through optical pumping and a microwave $\pi/2$-pulse, a test rotation $R_\mathrm{T}$ is applied, and the result $J_z = (N_\uparrow - N_\downarrow)/(N_\uparrow + N_\downarrow)$ is measured using the probe.  }
\label{fig:setup}
\end{figure}

The experimental system used for these studies was used to generate conditionally spin-squeezed states and is described in Ref.~\cite{CBS11}.  We use an ensemble of $N = 7 \times 10^5$ $^{87}$Rb atoms laser cooled and trapped in an optical lattice at 823~nm (Fig.~\ref{fig:setup}). The clock states $\ket{F=2, m_F=0} \equiv \ket{\uparrow}$ and $\ket{F = 1, m_F =0} \equiv \ket{\downarrow}$ constitute a pseudo spin-$\frac{1}{2}$ system/qubit with transition frequency $f_a = 6.834$~GHz.  All atoms are initially optically pumped into $\ket{\downarrow}$. Following that, a microwave $\pi/2$-pulse rotates the Bloch vector up to the equator, initializing the system for the experiments. To a very good approximation, the effects of microwave amplitude and phase inhomogeneity across the atomic ensemble may be neglected in our experiments. The intrinsic phase noise of the microwave LO, Agilent E8257D, is sufficiently low that we could use a small modulation amplitude $\beta$ in order to remain in the linear response regime and yet not be affected by phase noise of the LO source. Details on how phase noise from the microwave LO affects measurement signal to noise can be found in~\cite{Chen2012}.

An experiment typically consists of a test rotation $R_\mathrm{T}$ using phase modulated resonant microwaves coupling the two-level system with Rabi frequency $f_R=40.4$~kHz. After the rotation is completed, the Bloch vector projection $J_z = ( N_{\uparrow}-N_{\downarrow})/(N_\uparrow + N_\downarrow)$ is obtained using a cavity-aided nondemoliton measurement of the state populations $N_{\uparrow, \downarrow} =   \sum_{i=1}^{N} (1 \pm \langle \sigma^i_z \rangle )/2$, where $\sigma^i_z$ is the Pauli spin operator corresponding to spin $i$. The deflection $j_z$ is obtained from the measured $J_z$ and the $J^\circ_z$ for the same rotation without phase modulation using $j_z = J_z - J^\circ_z$.  For future reference, $\widetilde{j}_z$ corresponds to the special case of rotation axis about $\vc{\hat{x}}$, i.e. $\phi_R = 0$, and initial Bloch vector $\vc{J}_i = \vc{\hat{x}}$.

The nondemolition measurement is implemented by measuring the size of the collective vacuum Rabi splitting $\Omega_\uparrow$~\cite{Zhu90} generated by the coupling of a degenerate optical cavity mode to the $\ket{\uparrow} \to \ket{e}$ transition, where the optical excited state $\ket{e} \equiv \ket{F' = 1, m_F = 0}$. The size of the splitting $\Omega_\uparrow$ depends only on the total atomic population in $\ket{\uparrow}$ as $\Omega_\uparrow = \sqrt{N_\uparrow} 2g$, where $2g$ is the single atom vacuum Rabi frequency, a coupling constant determined by accurately known atomic properties and cavity geometry.  The size of the splitting is measured by sweeping a probe laser across the resonances and fitting the transmitted power to Lorentzians. The population $N_\uparrow$ is determined from the measured splitting using $N_\uparrow  = (\Omega_\uparrow/2g)^2$. Repeating the same procedure after a microwave $\pi$-pulse swaps the atomic populations determines the population $N_\downarrow$. From the measured populations $N_\uparrow$ and $N_\downarrow$, we obtain the quantity $J_z$.

\subsection{Response to Coherently Phase Modulated Rotation} 
\begin{figure}
\includegraphics[width=3.4in]{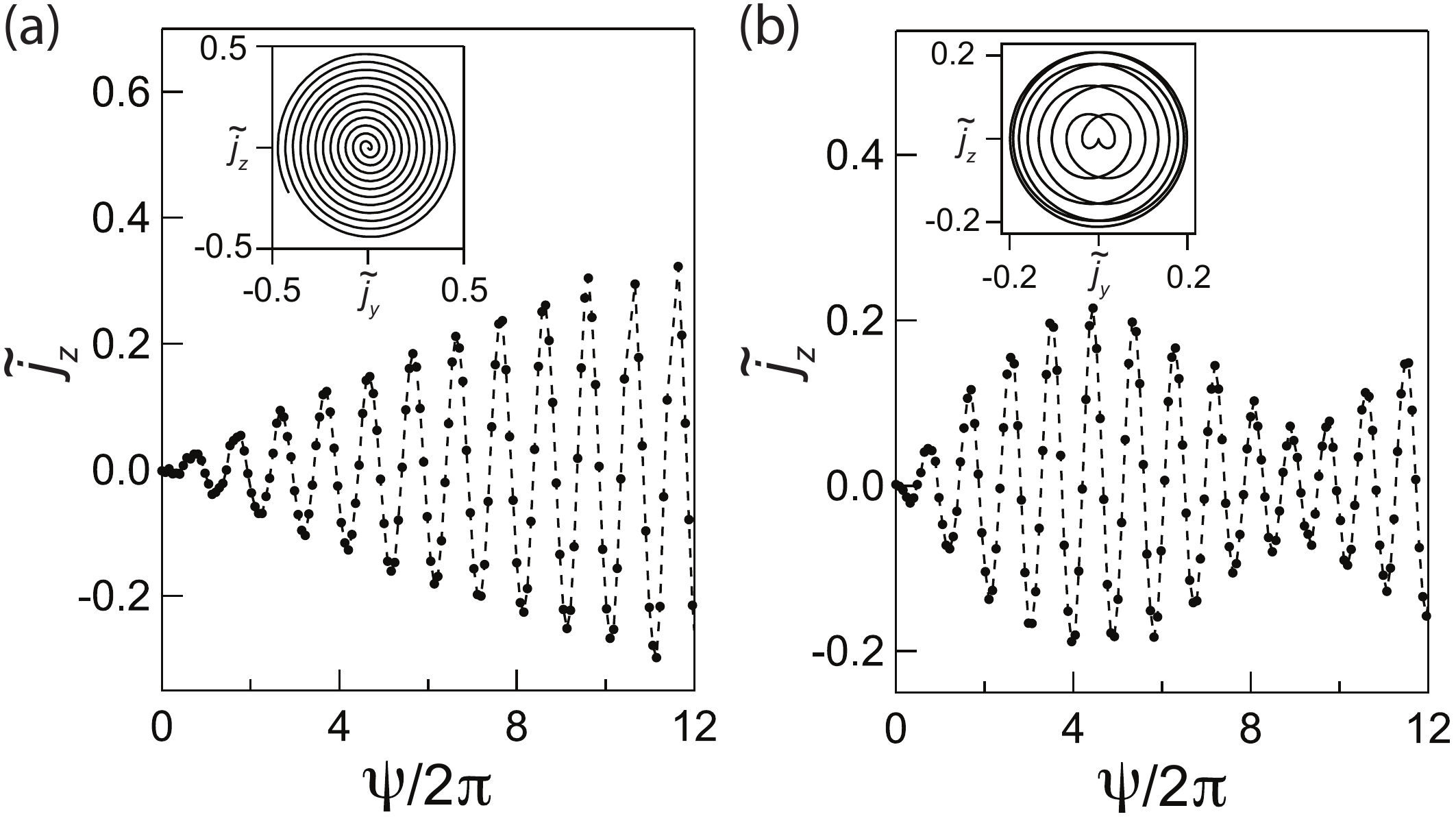}
\caption{\textbf{Phase Modulated Rotations.} Examples of the experimentally measured evolution of the Bloch vector along the measurement axis $\vc{\hat{z}}$ versus rotation angle $\psi$, or equivalently time, for (a) resonant modulation $f_m = f_R = 40.4~\mathrm{kHz}$, with $\beta = 0.0125$~rad, and (b) non-resonant modulation $f_m = 1.125 f_R$, with $\beta = 0.025$~rad. The modulation phase $\alpha_m$ is $0$ for both (a) and (b). Dashed lines joins the data points (solid circles) to help guide the eye. The system responds with frequency components at $f_m$ and $f_R$, leading to the observed amplitude modulation of the response in (b). Insets show the ideal theoretical spiral trajectories of the Bloch vector about $\vc{\hat{x}}$.}
\label{fig:TimeDomain}
\end{figure}

The harmonic oscillator-like response to a phase modulated rotation is experimentally demonstrated in Fig. \ref{fig:TimeDomain}. The Bloch vector is prepared along $\vc{\hat{x}}$, then rotated nominally about $\vc{\hat{x}}$ using microwaves whose phase is modulated at a fixed frequency $f_m$. After a variable rotation angle $\psi$, the projection $\widetilde{j}_z$ is measured.  For phase modulation near resonance $f_m \approx f_R$, the envelope of the modulation grows roughly linearly with $\psi$, whereas away from resonance $f_m \ne f_R$, frequency components at $f_R$ and $f_m$ beat against one another to create amplitude modulation.  The insets of Fig.~\ref{fig:TimeDomain} show the ideal theoretical spiral trajectories about $\vc{\hat{x}}$ for both cases.

\section{Noise in Bloch Vector due to Phase Noise in a Single Rotation}\label{sec:NoiseSingleRot}

\subsection{Covariance Transfer Matrix} \label{subsec:TransferMatrix}
Having obtained and experimentally demonstrated the response of the Bloch vector to a coherently phase modulated rotation in Sec.~\ref{sec:deflection}, the goal of this section is to define a covariance transfer matrix that will allow the computation of the variance of the final Bloch vector projection along any arbitrary axis $\vc{\hat{n}}$ due to a randomly phase modulated rotation caused by phase noise in the LO.

We begin with modelling phase noise at a single discrete frequency $f_0$ by allowing the modulation phase $\alpha_m$ and modulation amplitude $\beta$ to take on random values between realizations of the rotation. Statistical results are obtained via ensemble averaging over all possible realizations of the phase modulated rotation. The modulation phase and amplitude are fixed in a single realization but random from one realization to the next. By drawing the modulation phase $\alpha_m$ from a uniform distribution between $0$ and $2\pi$, and the modulation amplitude $\beta$ from a Gaussian distribution with zero mean and variance $\langle \beta^2 \rangle$, we model the SSB phase noise $\mathcal{L}(f_m) =  \langle \beta^2 \rangle \delta(f_m - f_0) /4$~(see Appendix~\ref{sec:AppendixB} for details). 

We define the covariance transfer matrix as the outer product $T(\phi_R, \vc{J}^\circ_f, \psi, f_m) \equiv 4 \langle \vc{j}_f \vc{j}^\intercal_f \rangle / \langle \beta^2 \rangle$,  where $\langle \cdot \rangle$ denotes a statistical average over $\alpha_m$ and $\beta$. The normalization of $T$  (labels suppressed) is chosen so that integrating $\vc{\hat{n}}^\intercal \cdot T \cdot \vc{\hat{n}}$ over the SSB phase noise $\mathcal{L}(f_m)$ of the LO yields the noise variance $\langle (\vc{j}_f \cdot \vc{\hat{n}})^2 \rangle$ of the final Bloch vector projected along a measurement axis $\vc{\hat{n}}$.   Using the solutions for $\widetilde{j}_y, \widetilde{j}_z$ in Eq.~\ref{eq:jy} and \ref{eq:jz}, we find the covariance transfer matrix $\widetilde{T}(\psi, f_m)$ for the special case $\phi_R=0$ and $\vc{J}_i = \vc{\hat{x}}$
\begin{widetext}
\begin{align}
\widetilde{T}(\psi, f_m) &= 
\begin{pmatrix}
0  &  0          &  0 \\
0  &  \widetilde{T}_{yy}  &  \widetilde{T}_{yz} \\
0  &  \widetilde{T}_{zy}  &  \widetilde{T}_{zz} \\
\end{pmatrix}  \, , \\
\widetilde{T}_{yy}(\psi, f_m)  &=  \frac{2}{(1-x^2)^2} \times \Bigl[ \left(\cos\psi - \cos\left(x \psi \right) \right)^2  +  \left( x \sin\psi - \sin\left(x \psi \right) \right)^2  \Bigr]  \, , \\
\widetilde{T}_{zz}(\psi, f_m)  &=  \frac{2}{(1-x^2)^2} \times \Bigl[ x^2 \left(\cos\psi - \cos\left(x \psi \right) \right)^2  +  \left( \sin\psi - x \sin\left(x \psi \right) \right)^2  \Bigr]  \, , \\
\widetilde{T}_{yz}(\psi, f_m) &= \widetilde{T}_{zy}(\psi, f_m)  =  \frac{2}{1-x^2}  \left(\cos\psi - \cos\left(x \psi \right) \right)   \sin\psi   \, .
\end{align}
\end{widetext}
The covariance transfer matrix $T$ for arbitrary rotation axis azimuthal angle $\phi_R$ and ideal final Bloch vector position $\vc{J}_f^\circ = (J_x^\circ, J_y^\circ, J_z^\circ)$ is derived using the small rotation $r$ specified in Eq.~\ref{eq:r(phi)} and \ref{eq:r(0)}. The full analytic expression for $T$ is cumbersome but can be conveniently obtained through the transformation  
\begin{equation}
T = D(\phi_R, \vc{J}_f^\circ) \widetilde{T}(\psi, f_m) D(\phi_R, \vc{J}_f^\circ)^\intercal \, , \label{eq:Ttransform}
\end{equation} where
\begin{equation}
D(\phi_R, \vc{J}_f^0) =
\begin{pmatrix}
0  &  -J_y^\circ  &  -J_z^\circ \cos \phi_R \\
0  &  J_x^\circ   &  -J_z^\circ \sin \phi_R \\
0  &  0          &  J_x^\circ \cos \phi_R + J_y^\circ \sin \phi_R
\end{pmatrix} \, . \label{eq:Dmatrix}
\end{equation}

\begin{figure}
\includegraphics[width=3.4in]{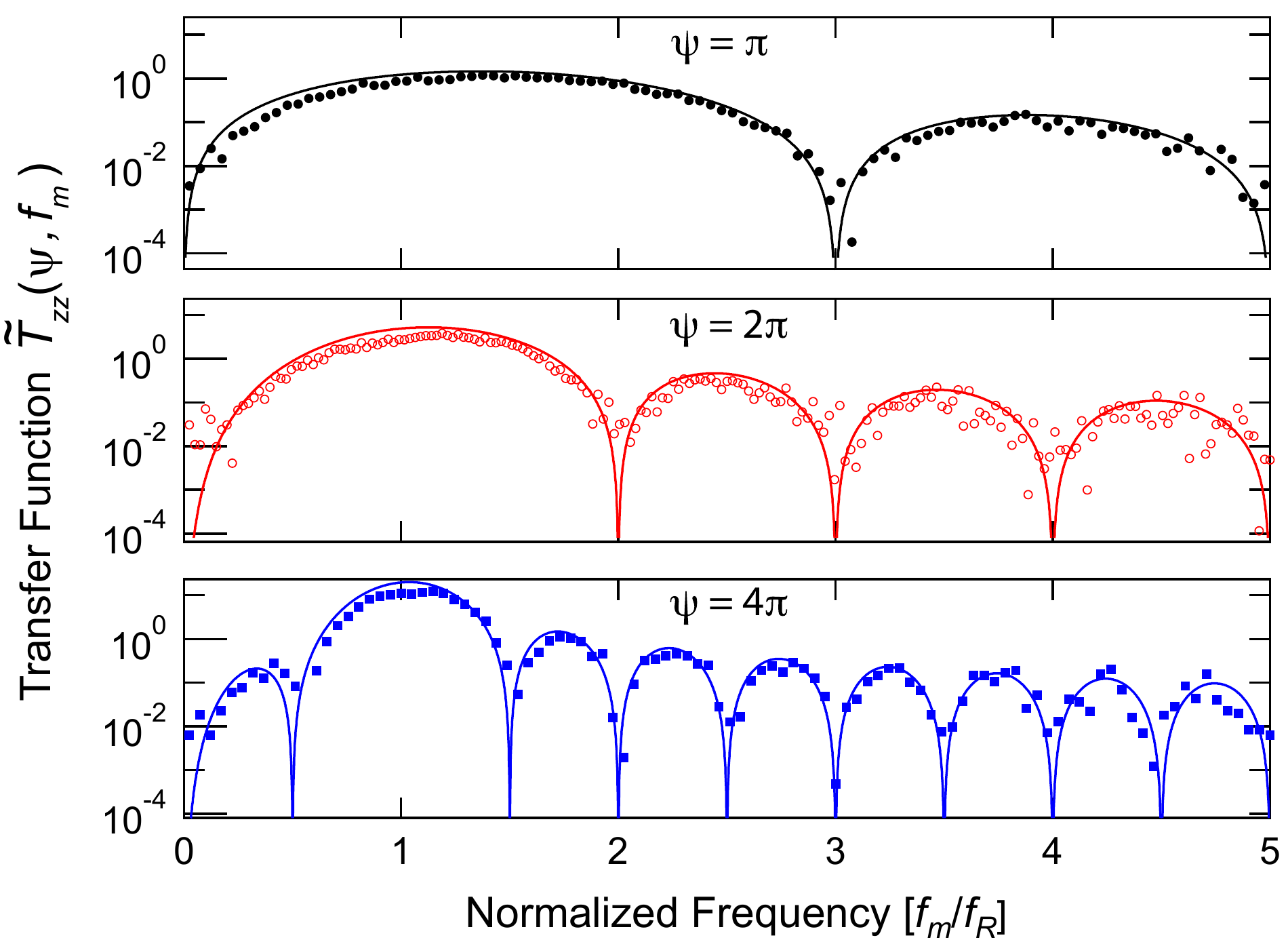}
\caption{(color online). \textbf{Transfer Functions.} Examples of experimentally measured transfer function  $\widetilde{T}_{zz}(\psi, f_m)$ for $\psi = \pi$ (black filled circles), $2\pi$ (red hollow circles) and $4\pi$ (blue filled squares) overlaid on parameter-free theoretical curves. The amplitude of the phase modulation was $\beta = 0.125, 0.05, 0.0625$~rad for the $\psi = \pi, 2\pi, 4\pi$ transfer functions respectively.  The modulation amplitude $\beta \ll 1$ is chosen to keep the response small to remain in the linear regime, and yet large enough to resolve the nulls at integer multiples of $f_R$. The transfer function does not depend on $\beta$ as long as $\beta \ll 1$ because both the response $\widetilde{j}_z$ and the modulation amplitude are proportional to $\beta$.} 
\label{fig:ExptTF}
\end{figure}

As an example of this general result, we experimentally measure the projection $\widetilde{j}_z$ after rotating $\vc{J}_i = \vc{\hat{x}}$ about $\vc{\hat{x}}$ through different angles $\psi = \pi$, $2\pi$ and $4\pi$ while phase modulating the microwave source.  Integer multiples of $\pi$ were chosen to minimize sensitivity to intrinsic phase noise of the microwave source near DC. The transfer function $\widetilde{T}_{zz}$ is obtained from averaging over four discrete values of the modulation phase $\alpha_m = \{0, \frac{\pi}{2}\, \pi, \frac{3\pi}{2}\}$, while keeping the amplitude $\beta$ constant, as $\widetilde{T}_{zz}(\psi, f_m) = \frac{1}{\beta^2}\sum^3_{n=0} \widetilde{j}_z^2(0, \psi, \beta, f_m, \frac{n \pi}{2})$.  The measured and theoretical transfer function $\widetilde{T}_{zz}$ are shown in Fig.~\ref{fig:ExptTF}.

\subsection{Covariance Noise Matrix}
In the linear response/small signal limit, the noise in the Bloch vector in some bandwidth is simply the integral of the noise variances due to phase noise at frequency $f_m$ over the relevant frequency bandwidth. Integrating the covariance transfer matrix $\widetilde{T}$ over the SSB phase noise $\mathcal{L}(f_m)$ of the LO yields the covariance noise matrix, defined for the largest possible bandwidth, 
\begin{equation}
\widetilde{V}(\psi) = \int^{\infty}_{0} \widetilde{T}(\psi, f_m)\mathcal{L}(f_m) \, d f_m \, .
\end{equation}
The covariance noise matrix transforms to arbitrary $\phi_R$ and $\vc{J}_f^\circ$ in the same manner as the covariance transfer matrix in Eq.~\ref{eq:Ttransform} via 
\begin{equation}
V = D(\phi_R, \vc{J}^\circ_f) \widetilde{V}(\psi) D(\phi_R, \vc{J}^\circ_f)^\intercal \, .
\end{equation}
Using the covariance noise matrix, the variance in the projection along $\vc{\hat{n}}$  may be obtained as
\begin{equation}
\langle (\vc{j}_f \cdot \vc{\hat{n}})^2 \rangle = \vc{\hat{n}}^\intercal \cdot V  \cdot \vc{\hat{n}}  \, .
\end{equation}

As a useful example of quantifying the noise mapping from the LO onto the Bloch vector, consider a white noise spectrum $\mathcal{L}(f_m) = \mathcal{L}_\circ$. Integrating the white phase noise spectrum over the covariance transfer matrix $\widetilde{T}(\psi, f_m)$ yields
\begin{equation}
\widetilde{V}(\psi) = \mathcal{L}_\circ  \widetilde{\mathrm{NEB}} \, ,
\end{equation}
where the noise equivalent bandwidth matrix is 
\begin{equation}
\widetilde{\mathrm{NEB}} = \pi f_R \, \mathrm{sgn}(\psi) 
\begin{pmatrix}
0  &  0  &  0 \\
0  &  \psi - \frac{1}{2} \sin 2\psi   &  -\sin^2 \psi \\
0  &  -\sin^2 \psi   &  \psi + \frac{1}{2} \sin 2\psi
\end{pmatrix} \, . \label{eq:NoiseMatrixwhitenoise}
\end{equation}
A corollary to Eq.~\ref{eq:NoiseMatrixwhitenoise} is that as $|\psi|$ increases, the covariance transfer matrix $\widetilde{T}$ becomes more and more sharply peaked around the Rabi frequency $f_R$. Therefore for large $|\psi| \gg 1$, most of the Bloch vector noise contribution comes from phase noise near the Rabi frequency.  As $|\psi| \to \infty$, the covariance transfer matrix $\widetilde{T}$ approaches a delta function at $f_R$
\begin{equation}
\widetilde{T}(\psi, f_m) \sim \pi f_R |\psi| \mathcal{L}_\circ \, \delta(f_m - f_R) \, \begin{pmatrix}
0  &  0  &  0 \\
0  &  1  &  0 \\
0  &  0  &  1
\end{pmatrix} \, .
\end{equation}

\begin{figure}
\includegraphics[width=3.4in]{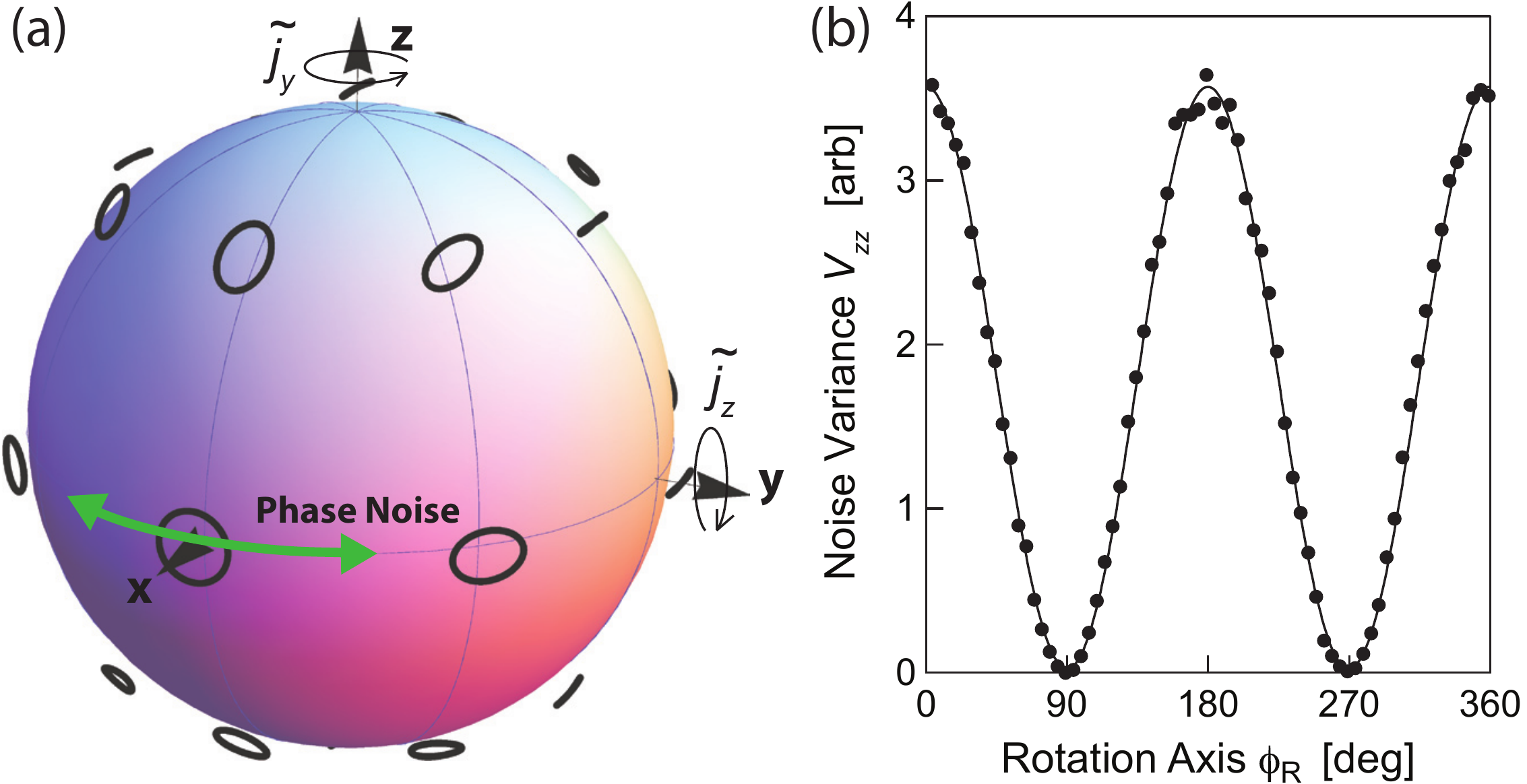} 
\caption{(color online). \textbf{Noise Mapping.} (a) Phase modulation of the LO causes modulation of the rotation axis (green arrow) about its mean orientation, here along $\vc{\hat{x}}$ or $\phi_R=0$.  The final Bloch sphere of points is deflected by an amount described by a small rigid rotation $r(0) = R_y(-\widetilde{j}_z) R_z(\widetilde{j}_y)$.  The deflection of a Bloch vector depends on the ideal final vector $\vc{J}_f^\circ$.  If a set of possible deflections for an ideal final Bloch vector $\vc{J}_f^\circ$ along $\vc{\hat{x}}$ is described by a circle  $\widetilde{j}_y^2 + \widetilde{j}_z^2 = \mathrm{const}$ (shown outside sphere for clarity), the same set of deflections for other $\vc{J}_f^\circ$ are described by ellipses and lines centered at $\vc{J}_f^\circ$.  (b) The noise mapping shown in (a) for $\phi_R = 0$ and for $\vc{J}_f^\circ$ on the equator can be equivalently demonstrated by keeping the final vector oriented along $\vc{J}_f^\circ=\vc{\hat{x}}$ and varying the rotation axis $\phi_R$.  The observed noise variance $V_{zz}$ (solid circles) of the vector projection along $\vc{\hat{z}}$ varies as the predicted  $\cos^2 \phi_R$ (solid line). The experiment was performed by applying phase modulation at a discrete frequency $f_m = 1.125 f_R$ to a rotation about $\vc{\hat{x}} \cos \phi_R + \vc{\hat{y}} \sin \phi_R$ that nominally rotates the Bloch vector $\vc{J}_i = \vc{\hat{x}} \cos 2\phi_R + \vc{\hat{y}} \sin 2\phi_R$ through an angle $\psi = \pi$ to $\vc{J}_f^\circ=\vc{\hat{x}}$. Averaging $j_z^2$ over the four modulation phases $\alpha_m = \{ 0, \frac{\pi}{2}, \pi, \frac{3\pi}{2} \}$ simulates phase noise at a single discrete frequency $f_m$.}
\label{fig:NoiseMapping}
\end{figure}

The noise mapping for any rotation axis in the $x-y$ plane can be obtained by transforming to arbitrary $\phi_R$, but keeping $\vc{J}_f^\circ = \vc{\hat{x}}$. We find the non-zero elements are $V_{zz} =\widetilde{V}_{zz} \cos^2\phi_R$, $V_{yy} =\widetilde{V}_{yy}$, and $V_{yz} =\widetilde{V}_{yz} \cos \phi_R$. Note that the variance $V_{zz}$ can be driven to zero by applying the rotation perpendicular to the Bloch vector as shown in Fig.~\ref{fig:NoiseMapping}(b). Alternately,  keeping $\phi_R=0$, but letting $\vc{J}_f^\circ = \vc{\hat{x}} \cos \theta + \vc{\hat{z}} \sin\theta$, i.e. the ideal final vectors lie in the $x-z$ plane, one finds $V_{zz, yy} =\widetilde{V}_{zz, yy} \cos^2\theta$ and $V_{xx} =\widetilde{V}_{zz} \sin^2 \theta$.  This noise mapping is graphically shown in Fig.~\ref{fig:NoiseMapping}(a).

\section{Noise in Bloch Vector due to White Phase Noise From Multiple Rotations}\label{sec:NoiseMultRot}

\subsection{Noise Propagation}\label{subsec:NoisePropagation}
The single-rotation covariance noise matrix $V$ allows us to analyze two crucial building blocks for coherent manipulation of quantum systems -- composite pulses, used to suppress static amplitude and detuning errors, and pulse sequences designed to reduce qubit decoherence, such as spin echo and dynamical decoupling type sequences. Constituent rotations in composite pulses are applied in a back-to-back manner, leaving as little time as possible between the rotations. In contrast, time separation beween rotations in a pulse sequence may be comparable or much longer than the time it takes execute a rotation. 

Assuming a white phase noise spectrum $\mathcal{L}(f_m) = \mathcal{L}_\circ$, noise from different rotations become statistically independent regardless of time separation between rotations. The time separation between rotations is, however, important when dephasing of the Bloch vector sets a timescale $\tau_c$ for the Bloch vector to phase diffuse by $\sim 1$~rad. In that case, it becomes important to ensure that the shortest time separation $\tau$ in the sequence satisfies $\tau \ll \tau_c$ so that the opening angle between the Bloch vector and rotation axis is well defined in order to apply the formulae presented below. This is a reasonable condition given that pulse sequences designed to reduce decoherence also  require $\tau \ll \tau_c$ in order to be effective. In this context, the formalism for treating noise from a composite pulse, and noise from a pulse sequence are the same.

We now present the noise propagation that gives the multiple-rotation covariance  noise matrix $W$. The symbol $W$ for the multiple-rotation covariance noise matrix is chosen to differentiate it from the single-rotation covariance noise matrix $V$.  A pulse sequence consists of $N$ rotations with the $k$th rotation given by 
\begin{equation}
R_k=R(\phi_k, \psi_k) \, . 
\end{equation}
In the absence of noise, the ideal Bloch vector after the $k$th rotation is 
\begin{equation}
\vc{J}_k^\circ = R_k \cdots R_1 \vc{J}_i \, .
\end{equation}
The added noise from just the $k$th rotation is 
\begin{equation}
V_k = D(\phi_k, \vc{J}^\circ_k) \widetilde{V}(\psi_k)  D(\phi_k, \vc{J}_k^\circ)^\intercal \, .
\end{equation}
By accounting for how the noise from the previous rotation $W_{k-1}$ is transformed by subsequent rotations, the total noise after the $k$th rotation may be computed iteratively using 
\begin{equation}
W_k = R_k W_{k-1} R^\intercal_k + V_k \, ,
\label{eq:NoiseProp}
\end{equation}
\noindent with the initial condition $W_1 = V_1$. Finally, the noise variance along any arbitrary projection axis $\vc{\hat{n}}$ after the $k$th rotation is given by 
\begin{equation}
\langle (\vc{j}_k \cdot \vc{\hat{n}})^2 \rangle = \vc{\hat{n}}^\intercal \cdot W_k  \cdot \vc{\hat{n}}  \, ,
\end{equation}
where $\vc{j}_k$ is the noise deflection after the $k$th rotation.

\subsection{Average Infidelity}
While matrix elements of the covariance noise matrix $W$ depend on pulse sequence specifics and the initial Bloch vector,  we present here a simple formula that evaluates the average quality of a pulse sequence using only general properties of the pulse sequence. Within the quantum control and computing community, the state infidelity~\cite{Jozsa94}
\begin{equation}
1 - F = \mathrm{Tr}(W)/4
\end{equation}
is an important measure of the rotation quality. While the state infidelity depends on pulse sequence details, the state infidelity averaged over the Bloch sphere of possible initial states
\begin{equation}
\langle 1-F \rangle = \Psi \pi f_R \mathcal{L}_\circ /3
\end{equation}
depends only on the total rotation angle $\Psi = \sum_{k=1}^N \psi_k$ and not the rotation axes $\phi_k$.  Thus, a pulse sequence with smaller $\Psi$ is preferred over one with larger $\Psi$ if one is mainly concerned with the average fidelity. In spin echo and dynamical decoupling schemes, suppression of environment-induced decoherence typically improves with the number of pulses. However, this comes at the expense of increasing the average phase-noise-induced decoherence. It is, therefore, necessary to strike a balance between reducing environment-induced decoherence and reducing phase-noise-induced decoherence.

\subsection{Composite $\pi$-pulse Comparisons}
Composite pulses, designed to suppress static amplitude and detuning errors, have been thoroughly analyzed in the literature with regards to the degree of error cancellation~\cite{TCS85, Wimperis94, CLJ03, BHC04, RLL09}. The influence of phase noise on the Bloch vector through composite pulses, however, has received little attention in literature.

\begin{figure}
\includegraphics[width=3.4in]{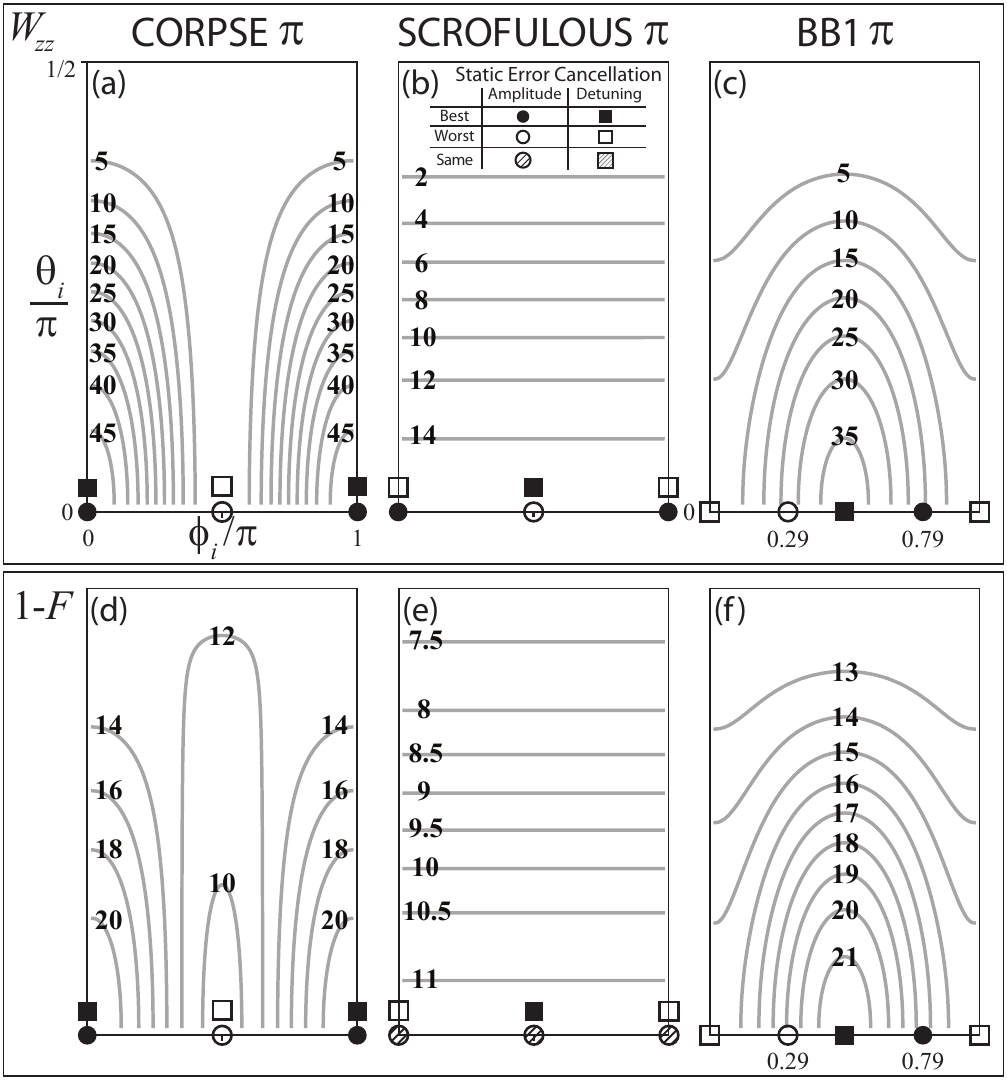}
\caption{\textbf{Composite $\pi$-pulses.} Variance $W_{zz}$ and infidelity $1-F$ of CORPSE~$\pi$-pulse (a, d), SCROFULOUS~$\pi$-pulse (b, e)  and BB1~$\pi$-pulse (c, f) versus initial Bloch vector orientation $(\theta_i, \phi_i)$. $\theta_i$ is measured from the $x-y$ plane and $\phi_i$ is measured from $\vc{\hat{x}}$. Contours levels are normalized to $f_R \mathcal{L}_\circ$. For scale, $\widetilde{W}_{zz}(\theta_i = 0, \phi_i = 0)$ for a simple $\pi$-pulse about $\vc{\hat{x}}$ is $\pi^2 f_R \mathcal{L}_\circ$. Points of best (solid circle), worst (hollow circle),  and same (hatched circle) order of static amplitude error cancellation, and points of best (solid square), worst (hollow square), and same (hatched square) order of static detuning error cancellation (offset vertically for clarity) are shown for Bloch vectors with $\theta_i =0$.  All plots have the same axes as (a). A quadrant of the Bloch sphere is shown here. The rest of the contour plot can be generated using reflection symmetry about the $x-y$ and $x-z$ plane. The rest of the static error cancellation points can be generated via $\phi_i \to \phi_i + \pi$. }
\label{fig:composite}
\end{figure}

Applying the noise propagation formalism presented in Sec.~\ref{subsec:NoisePropagation} to the commonly used composite $\pi$-pulse sequences: CORPSE, SCROFULOUS, and BB1~\cite{TCS85, Wimperis94, CLJ03}, which effectively implement a $\pi$-pulse about $\vc{\hat{x}}$, we summarize in Fig.~\ref{fig:composite} the final variance $W_{zz}$ and the state infidelity $1 - F$ versus the initial Bloch vector $\vc{J}_i$ specified by its polar angle $\theta_i$ and azimuthal angle $\phi_i$.  Expressions for the composite pulse rotation sequences, covariance noise matrices, and state infidelities are provided in Appendix~\ref{sec:AppendixC}.

The single quadrature variance $W_{zz}$ is of interest in metrology applications, particularly for manipulating spin-squeezed states, as added noise in the squeezed quadrature can potentially destroy the squeezing.  The state infidelity $1-F$, which includes variances from the two transverse spin components perpendicular to the ideal final Bloch vector, is particularly pertinent in quantum control for quantifying the overall quality of the rotations.

It is generally not possible to minimize phase noise sensitivity and optimize static error cancellation simultaneously. To understand the tradeoffs between phase noise sensitivity and static error cancellation, we compare and contrast the two for initial Bloch vectors in the $x-y$ plane, i.e. $\theta_i = 0$, leaving $\phi_i$ as the only degree of freedom. We use $W_{zz}$ and $1-F$ as the basis for evaluating sensitivity to phase noise.  

A static fractional amplitude error $\epsilon$ results in an error $\epsilon \psi$ in the rotation angle, and a static detuning error $\delta = (f_{LO} - f_a)/f_R$, where $f_{LO}$ is the LO  frequency, causes the rotation axis to be tilted up from the $x-y$ plane by an angle $\arctan (\delta)$. The degree to which the static errors $\epsilon, \delta$ are suppressed can be characterized by the static error squared $W_{zz, \mathrm{st}} = j_{z,\mathrm{st}}^2(\epsilon, \delta, \phi_i)$, and the infidelity due to static error  $1-F_{\mathrm{st}} = \mathrm{Tr}(W_{\mathrm{st}}(\epsilon, \delta, \phi_i))/4$. These definitions, in direct analogy to corresponding quantities for phase noise, allows meaningful comparison of the phase noise sensitivity and static error cancellation on the same footing in the next section.

\vspace{-.5cm}

\subsubsection{CORPSE $\pi$-pulse}
The CORPSE $\pi$-pulse, used to suppress static detuning error, has the best static error cancellation at $\phi_i = 0, \pi$ where both $W_{zz, \mathrm{st}}, 1-F_{\mathrm{st}} = O(\delta^6)$. However, it is also most sensitive to phase noise at $\phi_i=0, \pi$ (see Fig.~\ref{fig:composite}(a, d)).    At $\phi_i = \pi/2$, the impact of phase noise is minimized at the expense of static error cancellation as both $W_{zz, \mathrm{st}}, 1-F_{\mathrm{st}} = O(\delta^4) + O(\epsilon^2)$.

\subsubsection{SCROFULOUS $\pi$-pulse}
In contrast to the CORPSE $\pi$-pulse, sensitivity to phase noise does not vary with $\phi_i$ for the SCROFULOUS $\pi$-pulse designed to suppress static amplitude error.  As shown in Fig.~\ref{fig:composite}(b, e), one can simply choose $\phi_i$ to optimize the cancellation of static errors depending on the quantity $W_{zz, \mathrm{st}}$ or infidelity $1-F_{\mathrm{st}}$ to be optimized, and on the dominant source of static errors (amplitude or detuning) without altering the impact from phase noise.

\subsubsection{BB1 $\pi$-pulse}
Finally, the BB1 $\pi$-pulse, which compensates for amplitude error with little to no cost to the sensitivity to detuning error, has similar impact from phase noise where the order of amplitude error cancellation is best (worst) at $\phi_i \approx 0.79\pi \, (0.29\pi)$ as shown in Fig.~\ref{fig:composite}(c, f). Therefore one may choose to operate at $\phi_i \approx 0.79\pi$ where $W_{zz, \mathrm{st}} = O(\delta^2) + O(\epsilon^{10})$ and $1-F_{\mathrm{st}}  = O(\delta^2) + O(\epsilon^{8})$. In fact, the impact from phase noise is slightly lower at $\phi_i \approx 0.79 \pi$ compared to at $\phi_i \approx 0.29 \pi$. On the other hand, there is a tradeoff between suppressing  detuning errors and sensitivity to phase noise.  Detuning error cancellation is best at $\phi_i=\pi/2$ as both $W_{zz, \mathrm{st}},  1-F_{\mathrm{st}}$ do not scale with $\delta$ to any order if $\epsilon=0$. However, the impact of phase noise is also worst at $\phi_i = \pi/2$.

In general, careful evaluation of the relative scalings and contributions of phase noise, static amplitude and detuning errors is required to optimize the overall fidelity or specific noise quadratures.

\subsection{Spin Echo Pulse Sequences}
Spin echo and dynamical decoupling sequences constitute another class of manipulation protocols in quantum control and computing, important for suppressing qubit decoherence, or, for instance, to undo probe-induced dephasing~\cite{Appel09, Schleier-Smith10, LSV10, CBS11}. We analyze here spin echo sequences of the form $[ \tau - R(\phi_1, \pi)- 2\tau - R(\phi_2, \pi)- 2\tau \cdots - R(\phi_N, \pi)- \tau]$ using the formalism developed in Sec.~\ref{subsec:NoisePropagation} to find the covariance noise matrix $W$ for the sequence.

We consider the following two choices of rotation axes: (a) rotation axis always along $\vc{\hat{x}}$, or, (b) alternating between $\vc{\hat{x}}$ and $-\vc{\hat{x}}$ with the first $\pi$-pulse applied along $\vc{\hat{x}}$. Note that the widely used Carr-Purcell~\cite{CP54}, Carr-Purcell-Meiboom-Gill~\cite{MG58}, and Uhrig Dynamical Decoupling~\cite{Uhrig07} sequences are special cases of (a) corresponding to specific orientations of the Bloch vector with respect to the rotation axis $\vc{\hat{x}}$.

Writing the initial Bloch vector as $\vc{J}_i = (J_x^i, J_y^i, J_z^i)$, the covariance noise matrix for both choice (a) and (b) reads
\begin{equation}
W_N = N \pi^2 f_R \mathcal{L}_\circ
\begin{pmatrix}
1- J_x^{i \, 2}  & s_N J_x^i J_y^i &  s_N  J_x^i J_z^i \\
s_N J_x^i J_y^i  &  J_x^{i \, 2}  &  0 \\
s_N J_x^i J_z^i  &  0  &  J_x^{i \, 2} \\
\end{pmatrix} \, ,
\end{equation}
where $s_N = (-1)^{(N+1)}$. While the noise properties for the two choices are the same, their sensitivity to static errors are different. Choice (b) offers cancellation of static amplitude error as the Bloch vector nominally retraces its path while choice (a) accumulates static amplitude error as the Bloch vector keeps rotating about the same axis in the same sense. On the other hand, choice (a) does not accumulate static detuning error while choice (b) does. In future work, we will explore the possibility of engineering noise properties of spin echo pulse sequences.

\section{Conclusions} \label{sec:conclusion}
In summary, we have developed a general framework for analyzing the mapping of LO phase noise onto noise projections of a Bloch vector and extend the mapping to pulse sequences for the case of white LO phase noise. Detuning or transition frequency noise can be handled via the mapping $\beta \to \Delta/f_m$ where $\Delta$ is the frequency modulation amplitude. Results for special but important and illustrative cases are presented, which experimentalists can readily utilize for estimation or design. Future work will extend the analysis to non-resonant excitation, more complex spin echo or dynamical decoupling pulse sequences, and include the effects of $1/f$ and higher order phase noise, where time separation between pulses may no longer be ignored, and noise correlations between pulses play an important role.

\section{Acknowledgements}
We thank J.~Harlow and K.~Lehnert for loan of a microwave signal generator, and K.~Cox for helpful comments. This work was supported by the NSF AMO PFC and NIST. Z.C. acknowledges support from A*STAR Singapore, and J.G.B. acknowledges support from NSF GRF.

%Appendices
\renewcommand{\theequation}{A\arabic{section}.\arabic{equation}}
\renewcommand{\thefigure}{A\arabic{figure}}
\renewcommand{\thetable}{A\arabic{table}}

\appendix

\section{Phase Noise Definitions}\label{sec:AppendixA}
\setcounter{equation}{0}

The power spectral density of phase fluctuations $S_\phi(f_m)$ of an oscillator is the mean squared phase fluctuations $(\Delta \phi(f_m))^2$ at frequency offset $f_m$ from the carrier in a 1~Hz measurement bandwidth
\begin{equation}
S_\phi(f_m) \equiv \Delta \phi(f_m)^2 / \mathrm{Hz} \, .
\label{eq:phasenoisedef}
\end{equation}
The phase noise $S_\phi(f_m)$ has units of $\mathrm{rad}^2/\mathrm{Hz}$, and includes contributions from both upper and lower noise sidebands at $\pm f_m$.

One can measure $S_\phi(f_m)$ by mixing the oscillator-under-test with a reference oscillator at the same frequency, and with much lower phase noise. The phase of the reference oscillator is chosen so that the mixer output $v(t)$ is proportional to the relative phase difference $\phi(t)$ between the oscillator-under-test and the reference oscillator. The power spectral density $S_\phi(f_m)$ is computed from the autocorrelation function of $\phi(t)$ via the Wiener-Khinchin theorem as
\begin{equation}
S_\phi(f_m) = 2 \int_{-\infty}^{\infty} \langle \phi(t) \phi(t+\tau) \rangle_t  \, e^{-i 2\pi f_m \tau} \, d\tau \, ,
\end{equation}
where $f_m$ lies in the range $(0, \infty)$.

The definition of the single sideband (SSB) phase noise $\mathcal{L}(f_m)$ is
\begin{equation}
\mathcal{L}(f_m) \equiv \frac{1}{2} S_\phi(f_m) \, .
\label{eq:SSBdef}
\end{equation}
The units for $\mathcal{L}(f_m)$ are $\mathrm{rad}^2/\mathrm{Hz}$. It is also commonly expressed in the form $10 \log_{10} \mathcal{L}(f_m)$, which has units of dB below the carrier in a 1 Hz bandwidth (dBc/Hz). The mean squared phase fluctuations observed in a Fourier frequency range from $f_l$ to $f_h$ is given by
\begin{equation}
(\Delta \phi)^2 = 2 \int_{f_l}^{f_h} \mathcal{L}(f_m) \, df_m  \, .
\end{equation}

For an oscillator whose amplitude noise is much lower than its phase noise, the SSB phase noise $\mathcal{L}(f_m)$ is equivalent to the ratio of the power of a noise sideband $P_{\mathrm{SSB}}(f_m)$ at frequency offset $f_m$ in a 1~Hz measurement bandwidth to the power in the carrier $P_{\mathrm{car}}$ as measured on a radio/microwave/optical frequency spectrum analyzer
\begin{equation}
\mathcal{L}(f_m) = \frac{P_{\mathrm{SSB}}(f_m)}{P_{\mathrm{car}}} \, .
\end{equation}

\section{Relation between Mean Squared Modulation Amplitude and Single Sideband Phase Noise}\label{sec:AppendixB}
\setcounter{equation}{0}

We establish here the connection between the mean squared phase modulation amplitude $\langle \beta^2 \rangle$, used to normalize the covariance noise matrix $T$ in Sec.~\ref{subsec:TransferMatrix}, and the SSB phase noise $\mathcal{L}(f_m)$. We model the phase modulation $\phi(t) = \beta \sin(2\pi f_0 t + \alpha_m)$, at fixed modulation frequency $f_0$, as being drawn from a random  distribution of $\alpha_m$ and $\beta$. The modulation phase $\alpha_m$ is uniformly distributed from $0$ to $2\pi$, and the modulation amplitude $\beta$ is Gaussian distributed with zero mean and variance $\langle \beta^2 \rangle$. The time-averaged squared phase modulation should be further averaged over the distribution for $\alpha_m$ and $\beta$ to yield the statistical phase fluctuations as
\begin{equation}
\langle \langle \phi(t)^2 \rangle_t \rangle_{\alpha_m, \beta} = \langle \phi(t)^2 \rangle_{\alpha_m, \beta} = \frac{\langle \beta^2 \rangle}{2} \, ,
\label{eq:statavg}
\end{equation}
where we made use of the fact that averaging over time $t$ has the same effect as averaging over phase $\alpha_m$. Using Eq.~\ref{eq:phasenoisedef}, \ref{eq:SSBdef} and \ref{eq:statavg}, we obtain the relation between the SSB phase noise $\mathcal{L}(f_m)$ and mean squared modulation amplitude $\langle \beta^2 \rangle$ as
\begin{equation}
\mathcal{L}(f_m) = \frac{\langle \beta^2 \rangle}{4} \delta(f_m - f_0) \, .
\end{equation}

\medskip

\section{Composite and Single $\pi$-pulse: Covariance Noise Matrix and Infidelity}\label{sec:AppendixC}
\setcounter{equation}{0}

The covariance noise matrix $W$ and infidelity $1-F$ for the CORPSE $\pi$-pulse, SCROFULOUS $\pi$-pulse, BB1 $\pi$-pulse and a single $\pi$-pulse, which all effectively implement a $\pi$-pulse about $\vc{\hat{x}}$, are given below. The results assume a white LO phase noise spectrum $\mathcal{L}(f_m) = \mathcal{L}_\circ$. We specify the initial Bloch vector $\vc{J}_i$ by its polar angle $\theta_i$, measured from the $x-y$ plane, and its azimuthal angle $\phi_i$ measured from $\vc{\hat{x}}$ so that $\vc{J}_i = (\cos \theta_i \cos \phi_i, \cos \theta_i \sin \phi_i, \sin \theta_i)$.

\subsection{CORPSE $\pi$-pulse}
For the CORPSE $\pi$-pulse sequence $R\left(0, \frac{\pi}{3}\right)  R\left(\pi, \frac{5\pi}{3}\right)  R\left(0, \frac{7\pi}{3}\right)$  (time ordering right to left), used to suppress static detuning error, we have:
\begin{widetext}
\begin{align}
W_{xx} &=  \frac{1}{3} \pi f_R \mathcal{L}_\circ  \left(\left(13\pi - 3\sqrt{3}\right) \cos^2\theta_i \sin^2\phi_i +\left(13\pi + 3\sqrt{3}\right) \sin^2\theta_i\right)   \\
W_{yy} &=  \frac{1}{3} \pi f_R \mathcal{L}_\circ  \left(13\pi - 3\sqrt{3}\right) \cos^2\theta_i \cos^2\phi_i   \\
W_{zz} &= \frac{1}{3} \pi f_R \mathcal{L}_\circ  \left(13\pi+ 3\sqrt{3}\right) \cos^2\theta_i \cos^2\phi_i   \\
W_{xy} &= W_{yx} = \frac{1}{6} \pi f_R \mathcal{L}_\circ  \left(13 \pi - 3\sqrt{3}\right) \cos^2\theta_i \sin2\phi_i   \\
W_{yz} &=  W_{yz} = 0   \\
W_{xz} &= W_{zx} = \frac{1}{6} \pi f_R \mathcal{L}_\circ  \left(13\pi + 3\sqrt{3}\right) \sin2\theta_i \cos\phi_i  \\
1-F &= \frac{1}{12}  \pi f_R \mathcal{L}_\circ \left(13\pi -3\sqrt{3}\cos2\theta_i + \left(13\pi + 3\sqrt{3}\right) \cos^2\theta_i \cos^2\phi_i \right)
\end{align}
\end{widetext}

\subsection{SCROFULOUS $\pi$-pulse}
For the SCROFULOUS $\pi$-pulse sequence $R\left(\frac{\pi}{3}, \pi\right) R\left(\frac{5\pi}{3}, \pi\right) R\left(\frac{\pi}{3}, \pi\right)$ (time ordering right to left), used to suppress static amplitude error, we have:
\begin{align}
W_{xx} &= \frac{3}{2} \pi^2 f_R \mathcal{L}_\circ \left(2\cos^2\theta_i \sin^2\phi_i + \sin^2\theta_i \right)  \\
W_{yy} &= \frac{3}{2} \pi^2  f_R \mathcal{L}_\circ \left(2\cos^2\theta_i \cos^2\phi_i + \sin^2\theta_i \right)   \\
W_{zz} &= \frac{3}{2} \pi^2 f_R \mathcal{L}_\circ \cos^2\theta_i   \\
W_{xy} &= W_{yx} = \frac{3}{2} \pi^2 f_R \mathcal{L}_\circ  \cos^2\theta_i \sin2\phi_i    \\
W_{yz} &= W_{zy} = -\frac{3}{4} \pi^2 f_R \mathcal{L}_\circ \sin2\theta_i \sin\phi_i   \\
W_{xz} &= W_{zx} = \frac{3}{4} \pi^2 f_R \mathcal{L}_\circ \sin2\theta_i \cos\phi_i   \\
1-F &=  \frac{3}{16} \pi^2 f_R \mathcal{L}_\circ  (5+ \cos 2\theta_i)
\end{align}

\vspace{0.1cm}

\subsection{BB1 $\pi$-pulse}
The BB1 $\pi$-pulse sequence $R\left(0, \pi\right)  R\left(\phi_R, \pi\right) R\left(3\phi_R, 2\pi\right) R\left(\phi_R, \pi\right)$ (time ordering right to left),  where $\phi_R= \arccos\left(-\frac{1}{4}\right) \approx 104.5^\circ$, is used to compensate for amplitude error with little to no cost in the sensitivty to detuning error. The covariance noise matrix $W$ and infidelity $1-F$ for the BB1 $\pi$-pulse are:
\begin{align}
W_{xx} &= \frac{5}{4} \pi^2 f_R \mathcal{L}_\circ  \left(4\cos^2\theta_i \sin^2\phi_i + \sin^2\theta_i \right)   \\
W_{yy} &= \frac{5}{4} \pi^2 f_R \mathcal{L}_\circ  \left(4\cos^2\theta_i \cos^2\phi_i + 3\sin^2\theta_i \right)    \\
W_{zz} &= \frac{5}{4} \pi^2 f_R \mathcal{L}_\circ \cos^2\theta_i \left( 2  - \cos2\phi_i \right)  \\
W_{xy} &= W_{yx} = \frac{5}{2} \pi^2 f_R \mathcal{L}_\circ  \cos^2\theta_i \sin2\phi_i   \\
W_{yz} &=  W_{zy} = -\frac{15}{8} \pi^2  f_R \mathcal{L}_\circ \sin2\theta_i \sin\phi_i   \\
W_{xz} &= W_{zx} = \frac{5}{8} \pi^2 f_R \mathcal{L}_\circ \sin2\theta_i \cos\phi_i   \\
1-F &= \frac{5}{16} \pi^2 f_R \mathcal{L}_\circ \left(4 + 2\cos^2\theta_i -\cos^2\theta_i \cos2\phi_i \right) 
\end{align}

\subsection{Single $\pi$-pulse}
Finally, we provide expressions for the covariance noise matrix and infidelity for a single $\pi$-pulse $R\left(0,\pi\right)$ as a useful benchmark to compare against composite $\pi$-pulses. The noise covariance matrix $W$ and infidelity $1-F$ for a single $\pi$-pulse are:
\begin{align}
W_{xx} &=  \pi^2 f_R \mathcal{L}_\circ \left( 1 - \cos ^2\theta_i\cos ^2\phi_i \right)   \\
W_{yy} &=  \pi^2 f_R \mathcal{L}_\circ \cos ^2\theta_i \cos ^2\phi_i   \\
W_{zz} &= \pi^2 f_R \mathcal{L}_\circ \cos ^2\theta_i \cos ^2\phi_i    \\
W_{xy} &= W_{yx} = \frac{1}{2} \pi^2 f_R \mathcal{L}_\circ \cos ^2\theta_i \sin2\phi_i   \\
W_{yz} &= W_{yz} = 0   \\
W_{xz} &= W_{zx} = \frac{1}{2} \pi^2 f_R \mathcal{L}_\circ \sin2\theta_i \cos \phi_i   \\
1-F &= \frac{1}{4} \pi^2 f_R \mathcal{L}_\circ \left( 1 + \cos ^2\theta_i\cos ^2\phi_i \right) 
\end{align}

%\bibliography{PhaseNoise2012_bib}

%merlin.mbs apsrev4-1.bst 2010-07-25 4.21a (PWD, AO, DPC) hacked
%Control: key (0)
%Control: author (8) initials jnrlst
%Control: editor formatted (1) identically to author
%Control: production of article title (-1) disabled
%Control: page (0) single
%Control: year (1) truncated
%Control: production of eprint (0) enabled
%

\end{document}